\documentclass[conference]{IEEEtran}
\IEEEoverridecommandlockouts
\usepackage{cite}
\usepackage{amsmath,amssymb,amsfonts}
\usepackage{algorithmic}
\usepackage{graphicx}
\usepackage{textcomp}
\usepackage{xcolor}
\usepackage{float}
\usepackage[hidelinks]{hyperref}
\def\BibTeX{{\rm B\kern-.05em{\sc i\kern-.025em b}\kern-.08em
    T\kern-.1667em\lower.7ex\hbox{E}\kern-.125emX}}
\begin{document}

\title{Damping LFOs: Grid Following with Power Oscillation Damping vs. Grid Forming vs. PSS
}

\author{\IEEEauthorblockN{Tamojit Chakraborty}
\IEEEauthorblockA{\textit{Member, IEEE} \\ \textit{Siemens Energy} \\
\textit{tamojit91@gmail.com}\\
}
\and
\IEEEauthorblockN{Anamitra Pal}
\IEEEauthorblockA{Senior Member, IEEE \\ \textit{Arizona State University} \\
\textit{Anamitra.Pal@asu.edu}\\
}
\and
\IEEEauthorblockN{Sam Maleki}
\IEEEauthorblockA{Member, IEEE \\ \textit{RMS Energy Co. LLC} \\
\textit{sam.maaleki@gmail.com}\\
}
}
\maketitle

\begin{abstract}
Low-frequency oscillations (LFOs) present a significant challenge to the stability and reliability of power systems, especially in grids with a high penetration of renewable energy sources. Traditional grid-following (GFL) inverters have proven less effective in 
damping such oscillations. This paper presents a GFL-power plant controller with an auxiliary power oscillation damping control for damping LFOs. This approach is compared with a traditional power system stabilizer (PSS) for a two-area power system.
Next, the research is extended by deploying grid-forming (GFM) controls, which by actively controlling the voltage and frequency dynamics emulate the behavior of traditional synchronous generators.
The paper analyzes two GFM control strategies: virtual synchronous machine (VSM) and droop control, and demonstrates their effectiveness in damping LFOs in the test system. The simulation results reveal that the performance of the proposed GFM-VSM rivals that of the PSS and is better than the GFL-power oscillation damper.

\end{abstract}


\section{Introduction}
\par The increasing adoption of converter interfaced generation (CIG) by displacing conventional synchronous machines (SMs) has reduced the system’s inherent physical inertia and short circuit strength.
Declining inertia negatively impacts system stability with electromechanical oscillations, typically in the range of 0.2 Hz to 2 Hz, often being affected by the penetration of CIGs \cite{r1}.
A reduction in the damping of these oscillatory modes lowers the systems's stability margins resulting in sustained oscillations following an angle disturbance, heightening risks of unstable operating conditions.
The risks
vary based on grid conditions and the placement and control strategies of the CIGs 
\cite{r2}.



\par The process of damping low-frequency oscillations (LFOs) relates to small-signal stability, which refers to the ability of a network to maintain synchronization when subjected to minor disturbances. 
The small-signal stability concerns stem from
inadequate damping torque in SMs. 
Traditionally, power system stabilizers (PSSs) have been used 
to damp LFOs \cite{PAL2013638}.
The primary function of a PSS is to enhance the damping of generator rotor oscillations by regulating the excitation system through an external control variable input. Typically, PSS consist of a gain block, a washout filter, and phase compensation blocks to correct for the phase lag of the excitation system. 

\par The displacement of SMs and the rise of CIGs have intensified rotor oscillations and their impact on system frequency \cite{r8}. To address these issues, power oscillation damping (POD) controllers whose structures are similar to PSSs, have been used as auxiliary controls of flexible AC transmission system (FACTS) and high-voltage DC (HVDC) systems \cite{Pat1,9166750}. These devices dampen oscillations by injecting active power (POD-P) or reactive power (POD-Q) to regulate voltage at the point of interconnection. 

\par POD controllers used in conjunction with power plant controllers (PPCs) in grid-following (GFL) inverters
have been proposed for
actively adjusting the inverter’s output to counteract destabilizing oscillatory modes \cite{r6, r7}. A step forward is using grid forming (GFM) inverters \cite{b1}  that offer rapid response to disturbances in the network by utilizing a synchronization loop that emulates the behavior of a SM \cite{r3, r4, DR1}. For example,\cite{VSGNew} presents a centralized controller to damp LFOs utilizing virtual synchronous generators. 

\par Considering the status of present literature, this paper 
develops two new user-defined control models for damping LFOs in renewable-rich power systems and compares their performance for a multi-area test system.
Our main contributions are:
\begin{enumerate}
    \item \textit{Auxiliary POD function in PPC model}: A GFL-POD controller function is developed on a PPC user-defined model (UDM) that has two choices to damp LFOs: using active power, frequency (POD-P) or reactive power, bus voltage (POD-Q) signals. The structure of the POD consists of a deadband, a low pass filter, a washout block, and a lead-lag block for phase compensation.
    \item \textit{
    GFM controller Model}: A GFM UDM is developed in PSS/E with four control layers: the outer layer emulates a synchronous generator's swing dynamics with adjustable inertia and damping while a virtual excitation control provides a voltage reference. A current controller that utilizes voltage and frequency droop to generate current commands which forms a voltage source behind an internal impedance.
    \item \textit{Comparison of Strategies}: The developed POD and GFM UDMs are compared against different GFM strategies proposed in \cite{r5} and a traditional PSS for a two-area benchmark system.
    Detailed eigenvalue analysis results and damping ratios are presented for each strategy. 
\end{enumerate}

\section{Modeling and Implementation of Controller Strategies}

\subsection{Power Oscillation Damping Controller Architecture}

\par
\begin{figure*}[htb]
\centering
\includegraphics[width=7 in]{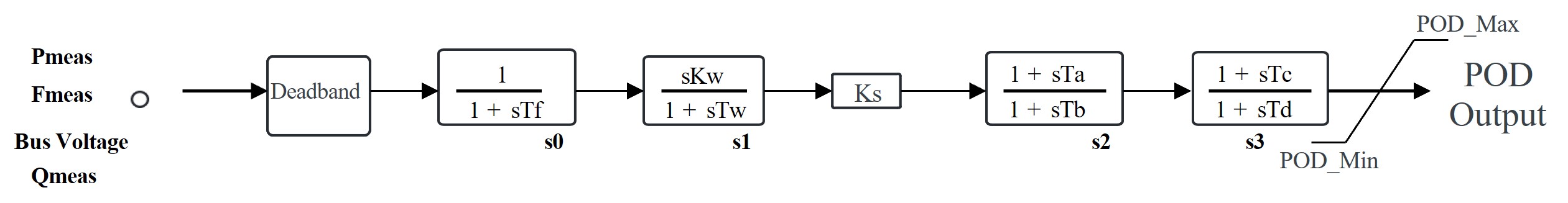}
\vspace{-1em}
\caption{Proposed GFL-POD Controller 
}
\label{fig:POD Control}
\end{figure*}

 This section presents the design and implementation of the GFL-POD controller.
Since the WECC generic REPCA model lacks POD functionality, a custom PPC model was developed in PSS/E. This PPC model offers several features \cite{b2}, including voltage, power factor, reactive power, and frequency control, in addition to the POD functionality. Note that the proposed POD control shares a structural resemblance to a conventional PSS. For POD-P, either active power measurement at a branch or bus frequency can be used, while for POD-Q, either reactive power measurement at a branch or bus voltage serves as the input. Fig. \ref{fig:POD Control} shows the block diagram of the POD controller. The dynamics of this controller are outlined 
below: 
\begin{equation}
\frac{ds_{0}}{dt}  = (1/T_{f})*[Vinp-s_{0}]
\label{Eq1}
\end{equation} 
\begin{equation}\frac{ds_{1}}{dt}  = (1/T_{w})*[K{w}*s0/T_{w}-s_{1}]
\label{Eq2}
\end{equation} 

\par The first block labeled $s_0$ is a low-pass filter that filters high-frequency signals, allowing only low-frequency components to pass through. It stabilizes the input signal (Vinp) by reducing noise and rapid fluctuations.
The washout block ($s_1$) filters out steady-state or low-frequency components, allowing only oscillatory (dynamic) components of the input signal to pass. This ensures that oscillations are damped effectively without impacting the steady-state power or voltage levels. The two lead-lag compensators ($s_2$ and $s_3$) are used to phase-shift the signal appropriately to counteract oscillations. The lead part introduces phase advance, which helps to counteract phase lags in the system, while the lag part stabilizes the response. These blocks adjust the phase and magnitude of the signal to maximize the damping of oscillations at specific frequencies. Finally, the limiters ensure that the output remains within a predefined range, preventing excessively high or low control signals that could destabilize the system. In summary, the POD controller uses a lag block for signal smoothing, a washout block to isolate oscillatory components, and lead-lag blocks to provide phase correction, allowing it to effectively dampen power oscillations in the system.

\subsection{Grid Forming Model Control Architecture}

\par The core control structure of a GFM comprises an outer voltage controller and an inner current controller. The outer control consists of active power-frequency (P-F) and reactive power-voltage (Q-V) control. In our proposed GFM model, the outer-level P controller has an integrator and a first-order time constant. They mimic the dynamics of the swing equation of a SM and generate a frequency reference.
The virtual excitation control generates a voltage reference using a proportional-integral (PI) controller. An inner current controller uses voltage and frequency droop to generate active and reactive current commands that are used to form a voltage source behind an internal impedance. Finally, a phase-locked loop (PLL) is used to convert the machine's rotational reference frame to the network's frame of reference. A PLL can still be utilized in a GFM, as this model supports rapid current injection (at a sub-transient timescale) by maintaining its internal voltage magnitude and angle. This conclusion is consistent with the findings in \cite{d2}.

\par
\begin{figure}[htb]
\centering
\includegraphics[width=3.45in]{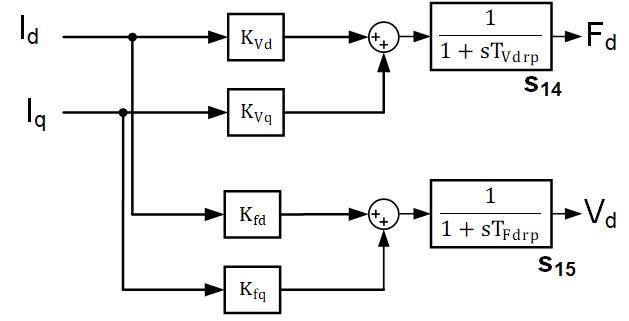}
\vspace{-2em}
\caption{ Droop Controller}
\label{fig: Droop Controller}
\end{figure}

\begin{figure}[htb]
\centering
\includegraphics[width=3.45in]{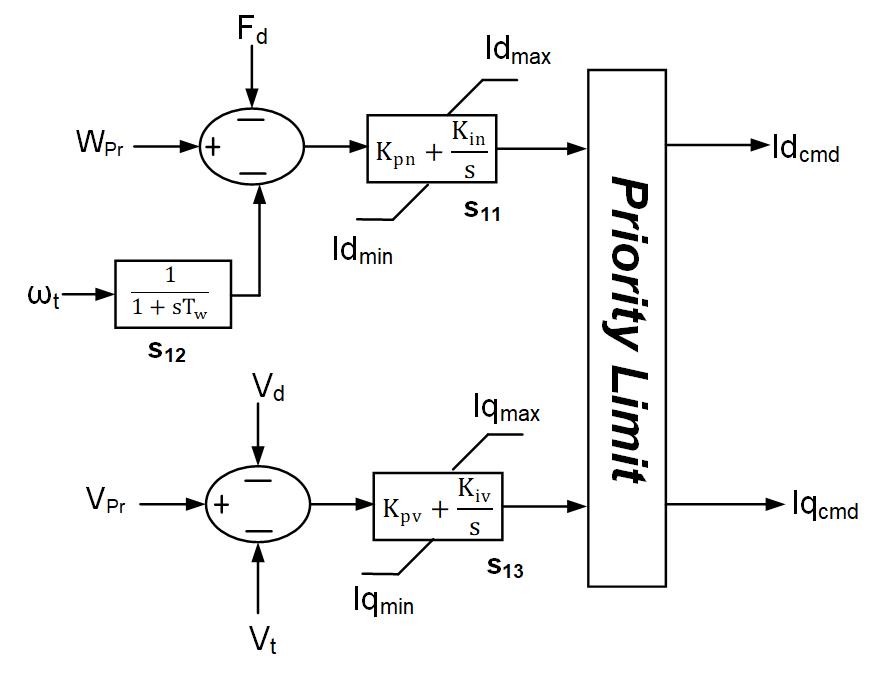}
\vspace{-2em}
\caption{ Current Controller}
\label{fig:Current Controller}
\end{figure}

\begin{figure*}[htb]
\centering
\includegraphics[width=7in]{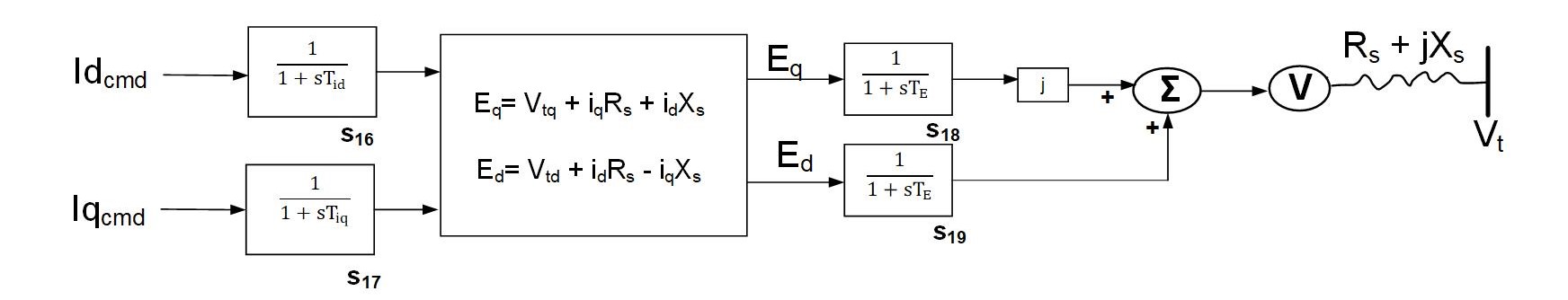}
\vspace{-1em}
\caption{VSC Interface for the Proposed GFM}
\label{fig:VSC_Interface}
\end{figure*}

 The GFM controller was developed as a custom UDM in PSS/E, utilizing FORTRAN code for implementation. PSS/E uses the modified Forward Euler method for numerical integration, updating state variables and their derivatives at each time step. In this model, current limits are implemented as per WECC MVS guidelines \cite{d1}.  Equations \eqref{Eq5} and \eqref{Eq6} represent the droop controller dynamics through simple first-order lag blocks (see Fig. \ref{fig: Droop Controller}), while current controller dynamics shown in Fig. \ref{fig:Current Controller} are outlined in \eqref{Eq7} and \eqref{Eq8}, respectively.  The voltage source interface of the GFM model is shown in Fig. \ref{fig:VSC_Interface}.
\begin{equation}
\frac{ds_{14}}{dt}  = (1/T_{vdrp})*[(I_{d}*K_{vd} + I_{q}*K_{vq}) - s_{14}]
\label{Eq5}
\end{equation} 
\begin{equation}
\frac{ds_{15}}{dt}  = (1/T_{fdrp})*[(I_{d}*K_{fd} + I_{q}*K_{fq}) - s_{15}]
\label{Eq6}
\end{equation} 
\begin{equation}
\frac{ds_{11}}{dt}  = K_{in}*[W_{pr}-F_{d}-s_{11}]
\label{Eq7}
\end{equation} 
\begin{equation}\frac{ds_{13}}{dt} = K_{iv}*[V_{pr}-V_{d}-s_{13}]
\label{Eq8}
\end{equation} 

\subsection{Other Grid Forming Control Strategies}
\par The WECC REGFMA1 \cite{r5}   model is employed to benchmark system performance against the proposed GFM and POD models. REGFMA1 incorporates P-F and Q-V droop controls, with active and reactive power limiting controls, and includes fault current limiting controls. The P-F control block participates in frequency regulation by adjusting the P output of the resource in response to grid frequency changes, while the Q-V control block aids in voltage regulation. 

\section{Stability Analysis}

\subsection{Two-Area System}\label{14b}
\par The control schemes presented in the previous section are employed to study the damping of LFOs in Kundur's two-area benchmark system. The system consists of two synchronous generators SG1 and SG2, connected through their corresponding step-up
transformers, and interconnected through two parallel lines between buses 2 and 3. A CIG device is also connected at Bus 2 through a step-up transformer. The SG1 and SG2 have a generation of 925 and 1300 MW, respectively, serving their local loads of 1000 and 1300 MW, respectively. The CIG device whose dynamic characteristics are varied to study different control strategies has a rating of 75 MW. It is considered that the impedance across the tie line connecting Buses 2 and 3 is variable. The loads have been modeled as ZI loads by assuming constant current for P and constant impedance for Q. 
SG1 and SG2 are modeled using GENROU mode with ESST4B excitation system and TGOV1 governor model. To evaluate the system performance with PSS, the CIG device is disconnected and the SG1 generates 1000 MW. The PSS model used in this study is the PSS2B.

\begin{figure}[htb]
\centering
\includegraphics[width=3.485 in]{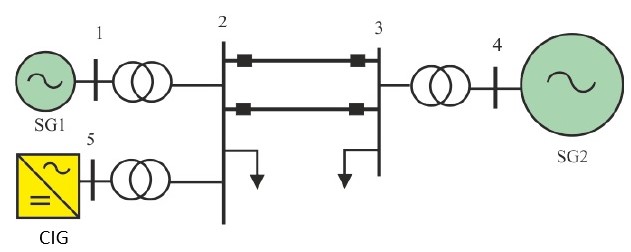}
\vspace{-2em}
\caption{Two-area System with CIG Device 
}
\label{fig:2area}
\end{figure}

\par The electrical model of the CIG with the POD used in this study consists of the generator converter model REGCA and electrical control model REECA and a user-defined PPC model in PSS/E. The variables VFlag and QFlag in the REECA model \cite{b3} are set to 1. The end-user has a choice through a flag to activate the POD in either the P control or the Q control mode. For POD-P, the user can either choose the frequency deviation at a bus or active power flow in a branch to modulate the direct axis current of the resource. A similar concept applies to the POD-Q logic where the end user can choose either a bus voltage or reactive power flow across a branch to modulate the quadrature axis current.  

\par For the GFM control strategies, the UDM described in the previous section is utilized. There are two possible model variants within the UDM: (a) virtual synchronous machine (VSM) combined with droop control, and (b) droop control alone. The VSM offers natural damping to the system, mimicking the dynamics of a SM. Similarly, the droop control-based GFM responds quickly to disturbances within the system. 
The ability of both the model variants were investigated in this study.
Lastly, the WECC REGFM model with typical parameters \cite{r5} is employed as an additional GFM control strategy.

\subsection{Simulation Set-up}\label{39a}

\begin{figure}[htb]
\centering
\includegraphics[width=0.485\textwidth]{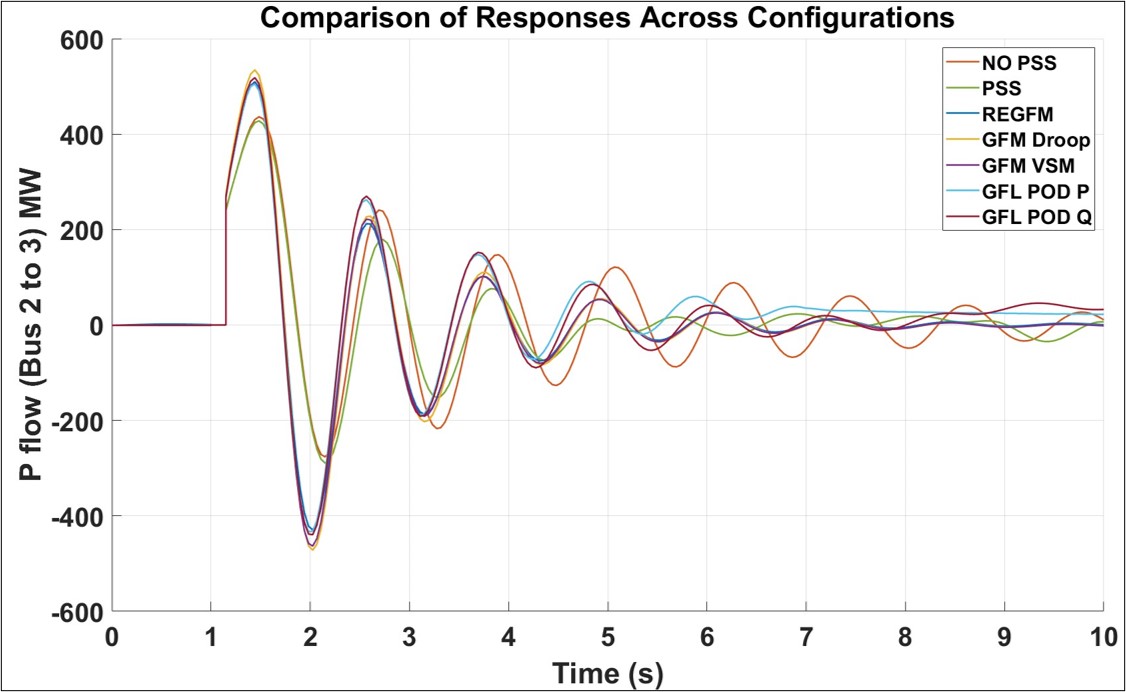}
\vspace{-1.5em}
\caption{Active Power Flow from Bus 2 to 3 when Tie Line Impedance Changes}
\label{fig:Power Plots}
\end{figure}

Several simulations have been performed on the two-area test system to gain insights into the performance of the different controller strategies described in this paper. To induce LFOs in the system the tie line impedance across Bus 2 and Bus 3 is varied 
to make the system reach a marginally stable condition.
Then,
a 15ms solid three-phase-to-ground is applied at Bus 2. As can be seen in Fig. \ref{fig:Power Plots}, the tie line shows LFOs.

\par Next,
a dual input PSS is applied at Bus 2 that uses generator active power and rotor speed deviation as inputs to enhance the damping of LFOs and improve system stability. By processing both inputs, it captures dynamic fluctuations in power flow and system frequency deviations, which indicate disturbances or imbalances. Each input passes through washout filters to remove steady-state components and lead-lag compensators to align the PSS’s corrective actions with oscillation frequencies. The combined output is applied to the generator's excitation system to counteract oscillations, providing robust damping and improved overall system stability.   

\par For the GFL case with POD as well as the GFM case,
a similar disturbance is introduced into the system, and each control strategy tries to mitigate the resulting oscillations through its dynamic response. In the GFL case, the POD-P and POD-Q controllers activate upon detecting LFOs, while the GFM units address system disturbances by closely regulating voltage magnitude and angle at their terminals.

\subsection{Eigenvalue Analysis}\label{39b}

\par
\begin{figure}[htb]
\centering
\includegraphics[width=3.485 in]{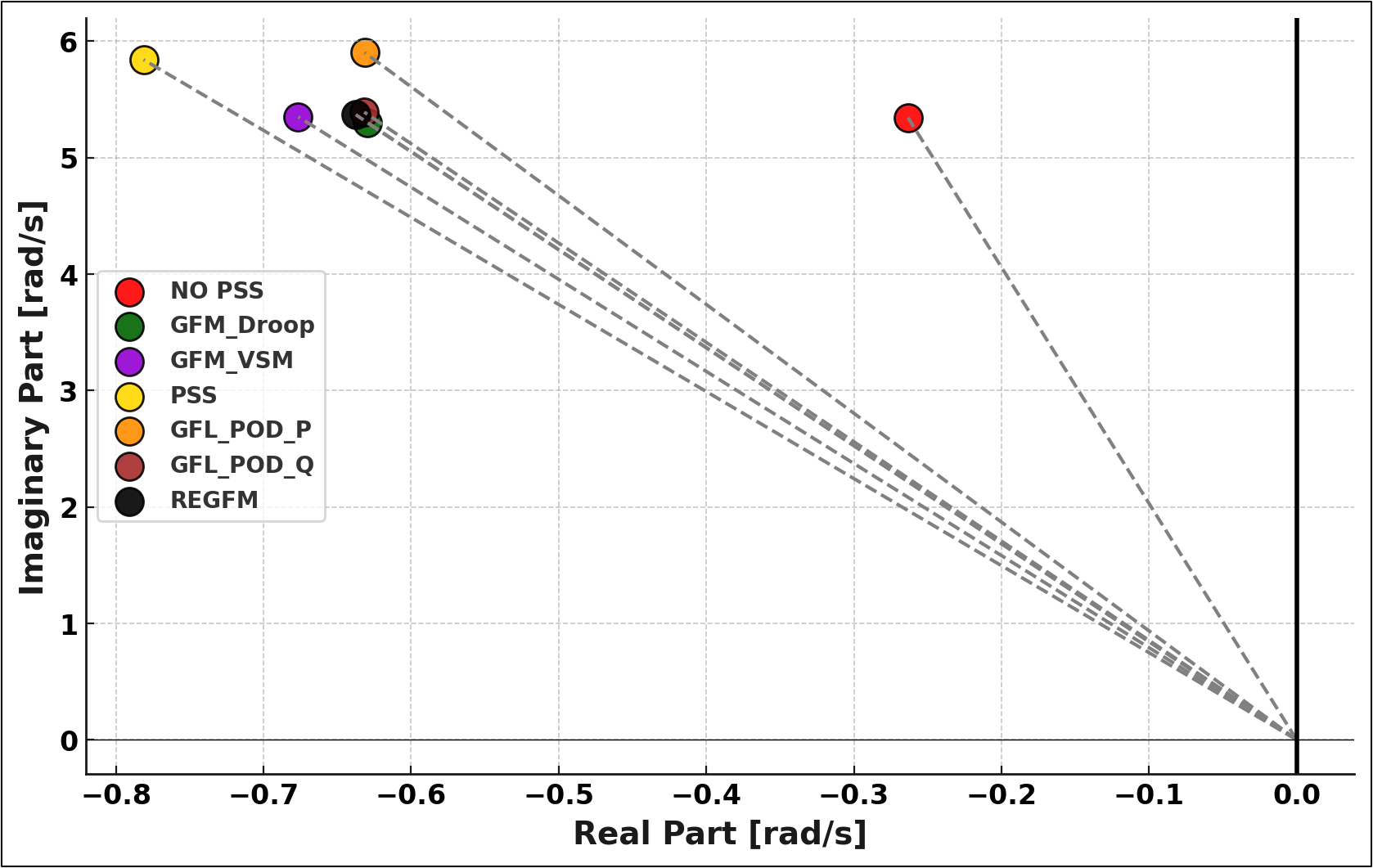}
\caption{Eigenvalues for Different Scenarios}
\label{fig:Eigen Values}
\end{figure}

 Fig. \ref{fig:Eigen Values} shows the eigenvalues for different configurations on the complex plane, with the X-axis representing the real part (damping) and the Y-axis representing the imaginary part (oscillation frequency). Negative real values suggest that oscillations decay over time, enhancing stability. Eigenvalues closer to the origin indicate lower frequencies and slower oscillations, while those farther away may indicate more dynamic responses.
Table \ref{tab:ranked_eigenvalues_damping} shows the eigenvalues for each control strategy. The eigenvalues are obtained by performing Prony Analysis using PSS/E built-in program PSSPLT.

\textit{Overall Comparison}: The eigenvalue comparison reveals that the PSS provides the highest damping effect overall, with the most negative real part, enabling oscillations to decay quickly and enhancing stability. The GFM-VSM also offers strong damping, though slightly less than the PSS, maintaining stability without significantly increasing oscillation frequency. The GFL equipped with POD-P provides moderate damping, better than NO PSS but less than 
PSS, GFM-VSM, GFM-Droop, REGFM, and POD-Q, with a relatively high frequency, suggesting a more dynamic response. In contrast, NO PSS has the lowest damping, as indicated by its least negative real part, which could lead to prolonged oscillations.

\textit{GFL-POD-P vs. GFL POD-Q}: The GFL with POD-P and POD-Q configurations show similar oscillation frequencies, as indicated by their close imaginary parts ($5.90218$ and $5.39$, respectively, in Table \ref{tab:ranked_eigenvalues_damping}). However, they differ in damping effectiveness. The POD-P has a real part of $-0.631503$ and a damping ratio of $10.6$\%, providing moderate damping. In contrast, the POD-Q has a slightly more negative real part ($-0.631767$) and a higher damping ratio of $11.6$\%, suggesting that it will stabilize oscillations slightly faster than POD-P,
making it marginally more effective in damping oscillations. 

\textit{GFM-VSM vs. GFM-Droop}: The GFM-VSM has a real part of $-0.676476$ and a damping ratio of $12.5$\%, indicating a strong damping effect and stable response. 
The GFM-Droop, with a real part of $-0.629683$ and a damping ratio of 11.8\%, offers slightly less damping, meaning oscillations decay a bit slower than with the GFM-VSM. Its imaginary part ($5.29911$) is also slightly lower, indicating a 
slightly slower oscillatory response. Overall, GFM-VSM provides a stronger damping effect and faster stabilization compared to GFM-Droop. 

\textit{Proposed GFM Control vs. Proposed GFL-POD-Q vs. REGFM} \cite{r5}: The GFM-VSM provides stronger damping and faster stabilization than REGFM, with a higher damping ratio of 12.5\% and more negative real parts for quicker oscillation decay. In contrast, the GFM-Droop and REGFM show similar performance, indicated by comparable damping ratios and eigenvalue real parts. Comparing the GFL against the GFM strategies, the GFM-VSM provides stronger damping and faster stabilization than GFL with POD-Q, though both show similar oscillation frequencies and dynamic behavior. Overall, the proposed GFM-VSM outperforms the others in terms of damping effectiveness and stability.






\begin{table}[ht]
\centering
\caption{Ranked Eigenvalues and Damping Ratios for Different Control Strategies}
\label{tab:ranked_eigenvalues_damping}
\begin{tabular}{|l|c|c|c|}
\hline
\textbf{Strategies} & \textbf{Real} & \textbf{Imag} & \textbf{Damping Ratio} \\
\hline
PSS & -0.781214 & 5.84032 & 13.3 \\
\hline
GFM\_VSM & -0.676476 & 5.34967 & 12.5 \\
\hline
REGFM \cite{r5} & -0.637492 & 5.36776 & 11.8 \\ \hline
GFM\_Droop & -0.629683 & 5.29911 & 11.8 \\
\hline
GFL\_POD\_Q & -0.631767 & 5.39000 & 11.6 \\
\hline
GFL\_POD\_P & -0.631503 & 5.90218 & 10.6 \\
\hline
NO PSS & -0.263220 & 5.34158 & 4.9 \\
\hline
\end{tabular}
\end{table}

\subsection{Dynamic Response to Load Change}\label{39c}

\par
\begin{figure}[htb]
\centering
\includegraphics[width=2.8 in]{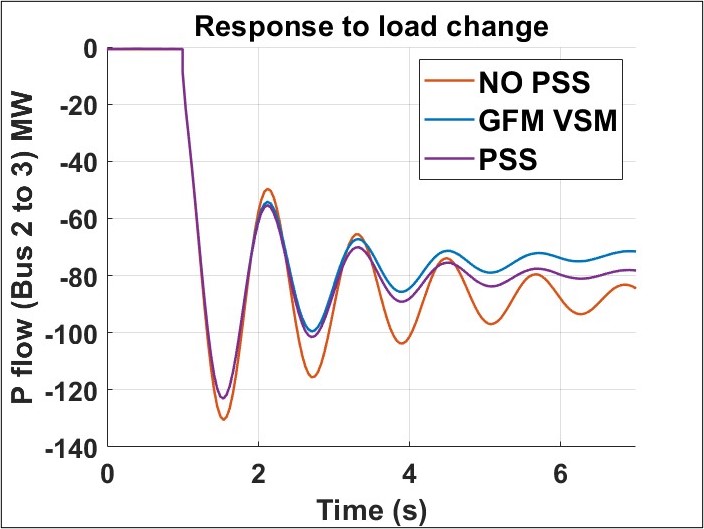}
\vspace{-0.5em}
\caption{Active Power Flow from Bus 2 to 3 when Load changes 
}
\label{fig:Load Rej}
\end{figure}

 This section analyzes the dynamic response of the two-area power system when subjected to a 150 MW load reduction at Bus 2. Without any PSS, the sudden load decrease causes oscillations in the tie line leading to prolonged instability (see Fig. \ref{fig:Load Rej}). With the GFM-VSM, the system’s response improves substantially (rivaling that of PSS). Specifically, the GFM-VSM immediately counteracts the oscillations in the tie line by modulating its power output. This rapid response highlights the GFM-VSM’s capacity to mimic the behavior of traditional SMs,
making them a valuable asset for enhancing grid resilience, especially under sudden changes in load or generation.
In summary, the results demonstrate that the proposed GFM-VSM with its machine-like dynamic performance, is well-suited for modern grids seeking enhanced stability and robust response for damping LFOs.

\section{Conclusion}

\par This paper examines the impact of LFOs on system stability and explores mitigation strategies amid the growing integration of CIGs and the retirement of traditional SMs. A GFL-POD controller is designed and implemented as an auxiliary function within a PPC, and a PSS/E UDM is developed to evaluate the controller's effectiveness in two modes: POD-P and POD-Q. Additionally, a GFM controller is developed to compare its functionality with the GFL-POD controls. The proposed GFM controller includes an outer loop that replicates SM swing dynamics, along with P-F and Q-V droop control. 
Both the GFL-POD and GFM controls are tested in a two-area test system, 
and a detailed small-signal stability analysis is conducted.
The results demonstrate the effectiveness of GFM controls (particularly, proposed GFM-VSM) in damping LFOs to a similar extent as a traditional PSS. This study provides valuable insights for a specific test system; future work will explore broader applicability in more complex networks.

\bibliographystyle{IEEEtran}
\bibliography{bibref}

\end{document}